\documentclass[preprint]{aastex}

\newcommand\simlt{\lower.5ex\hbox{$\; \buildrel < \over \sim \;$}}

\begin{document}

\title{Gravitational radiation from long gamma-ray bursts}
\author{Maurice H.P.M. van Putten}
\affil{Massachusetts Institute of Technology, Cambridge, MA 02139-4307} 

\begin{abstract}  
  Long gamma-ray bursts (GRBs) are probably powered by high-angular momentum
  black hole-torus systems in suspended accretion. The torus will radiate
  gravitational waves as non-axisymmetric instabilities develop.  The luminosity
  in gravitational-wave emissions is expected to compare favorably with the observed 
  istropic equivalent luminosity in GRB-afterglow emissions.
  This predicts that long GRBs are potentially the most powerful LIGO/VIRGO
  burst-sources in the Universe. Their frequency-dynamics is characterized by 
  a horizontal branch in the $\dot{f}(f)$-diagram.
\end{abstract}


\section{Introduction}

Cosmological gamma-ray bursts are the most relativistic events in the Universe,
at an observed rate of about two per day. The BATSE catalogue shows a bi-modal
distribution of GRB durations, with short bursts of about a second and long
bursts of about one minute \citep{kou93,pac99}. These events are probably associated 
with the formation of high-angular momentum low-mass black hole-torus systems
representing hypernovae \citep{woo93,pac97,pac98,bro00} in star-forming regions 
\citep{pac98,blo00}.

General arguments suggest that the rotational energy in a black hole can 
be similarly important as accretion, provided there is a channel in place 
to tap this energy. 
A maximally rotating black hole of mass $M\sim7M_\odot$ 
contains about $2M_\odot$ in rotational energy. Raleigh's criterion 
(e.g., \cite{kip90}) 
reminds us that rotating systems have a tendency to shed off angular momentum to 
larger distances, and a rotating black hole is no exception. This may be seen from 
the first law of black hole thermodynamics. For a black hole of mass $M$, angular 
momentum $J_H$ and angular velocity $\Omega_H$ we have \citep{bar73,haw75,haw76}:
$\delta M=\Omega_H\delta J_H+T_H\delta S_H$,
where $T_H$ the horizon temperature and $S_H$ ($\delta S_H\ge0)$ the entropy. Thus,
the specific angular momentum $a_p=\delta J_H/\delta M$ 
of a radiated particle exceeds that of the black hole:
$a_p\ge1/\Omega_H\ge 2M> M\ge J_H/M=a.$
Angular momentum is stored at a lower cost in radiation than 
in the black hole, and is emitted with an efficiency of at most
50\%. In vacuum, the decay of a Kerr black hole into Schwarzschild 
black hole is safeguarded by an angular momentum barrier, and
spontaneous emission of particles is exponentially small
\citep{teu73,pre73,teu74}.
Yet, black holes formed in hypernovae
will be surrounded by magnetized matter, in the form of an
accretion disk or torus. A similar state is expected 
from tidal break-up of a neutron star by a Kerr black hole
\citep{pac91,mvp99b}.

Ultrarelativistic leptonic outflows which form the input to GBRs may
be powered by black hole-spin in the presence of magnetic fields
\citep{mvp00a,mvp00b,hey00}. 
Long/short GRBs can hereby be 
associated with magnetic regulated 
suspended/hyper-accretion onto rapidly/slowly spinning black holes
\citep{mvp00c}. For long bursts, the suspended accretion state lasts
for the duration of spin-down of the black hole, whereby
the surrounding magnetized matter receives a powerful
torque $T=-\dot{J}_H$ from the angular momentum $J_H$ of the black hole
\citep{mvp99a,mvp99b,bro01} which arrests the inflow.
The black hole performs approximately isotropic work in
powering both the outflow and the Maxwell stresses onto the torus. 
This suggests to consider the possibility that the 
torus re-radiates this output, 
in part, in gravitational-wave emissions as it develops 
non-axisymmetric instabilities.

Here, we show that gravitational wave-emissions are expected in suspended 
accretion as a collateral feature to long GRB-afterglow emissions.
This feature is in many ways similar to new-born pulsars, 
which are well-known to radiate predominantly in gravitational waves \citep{sha83}.

\section{Torus in suspended accretion surrounding a rapidly spinning black hole}

The approximately isotropic work performed by the black hole takes
place over interconnecting magnetic field-lines regulated by the magnetic moment 
of the black hole in equilibrium with the torus magnetosphere
\citep{wal74,dok87,mvp00b,lee01}.
These field-lines comprise a torus magnetosphere supported by
surrounding baryonic matter and open field-lines to infinity,
as schematically indicated in Figure 1. The latter field-lines 
are endowed with conjugate radiative-radiative boundary conditions, 
whereas the former have radiative-Dirichlet boundary conditions. These different 
boundary conditions introduce interactions which will differ in details, 
e.g.: leptonic outflow to infinity and Maxwell stresses onto the torus, 
respectively. Nevertheless, the work performed per solid angle into these 
different field-lines is expected to be rather similar.

The magnetic connection of the black hole to the inner face of the
mediates Maxwell stresses by equivalence in poloidal topology to pulsar 
magnetospheres. An aligned rotator radiates Maxwell stresses
to infinity according to $-\dot{J}_{p}=\Omega_{p} A_p^2$,
where the subscript $p$ refers to the appropriate pulsar values
\citep{gol69}.
Analogously, causal Maxwell stresses are set up between the black hole 
and the inner face of the torus according to 
(adapted from \cite{tho86,mvp00c})
\begin{eqnarray}
\tau_+=(\Omega_H-\Omega_T)f_H^2A^2,
\end{eqnarray}
where $\Omega_T$ denotes the angular velocity of the torus and its similarly
shaped force-free magnetosphere,
and $2\pi f_HA$ denotes the flux in interconnecting 
magnetic field-lines - $2\pi A$ representing the net magnetic flux 
produced in the torus. This torque received from the black hole 
compensates for angular momentum losses in magnetic winds and radiation, which
arrests the inflow and enables a state of suspended accretion around a rapidly 
spinning black hole. The suspended accretion state is expected to be stable
on average \citep{mvp00c}.

There powerful shear between the inner and the outer faces of the torus
is, to leading order, dictated by Keplerian motion.
Some deviation away from Keplerian motion is expected due to the
presence of competing torques. They tends to bring the two faces in state
of super- and sub-Keplerian motion, with positive
radial pressure which promotes a slender shape.
The interface separating the two faces is expected to be
unstable, which favors turbulent mixing into a state of uniform 
specific energy across the torus. Mixing enhances differential 
rotation, as may be illustrated in the Newtonian limit, which
gives rise to the angular velocity $\Omega(r)\approx\Omega_K(1-(r-a)/a)^{1/2}$
as a function of radius $r$ for a torus of major radius $a$. Compression into a more slender
shape tends to reduce differential rotation. The net result  
should be that the characteristically Keplerian decrease 
of angular velocity with radius is approximately
preserved. The inner and other faces will have, respectively, angular velocities
$\Omega_\pm\approx \Omega_K\left(1\pm 3b/4a\right),$
where $b$ denotes their radial separation.
The same trend should hold in the Kerr metric.

\section{Gravitational radiation in suspended accretion}

Gravitational radiation from a torus surrounding a black hole tends to dominate
radio waves of the same frequency. This is generally due to the compact size in
the presence of gravitationally weak magnetic fields. Consider a torus
with ellipticity $\epsilon$, a magnetic moment $\mu_T$ and mass $m$ in rotation
about its center of mass. Its quadrupolar moments in magnetic 
moment and mass are, respectively, $\epsilon \mu$ and $\epsilon m$, which produce
luminosities (adapted from \cite{sha83}):
${\cal L}_{em}\approx\pi^{-1}(\Omega_T M)^4(\mu_T/M^2)^2\epsilon^2$
and 
${\cal L}_{gw}\approx ({32}/{5})(\Omega_T M)^{10/3} (m/M)^2\epsilon^2$
in geometrical units.
These emissions may be compared with, respectively, the luminosity in radio emission
$\sim \Omega^4_p\mu_p^2/\pi$ from an orthogonal pulsar and
in gravitational-wave emissions 
$\sim ({32}/{5})(\Omega_{orb} {\cal M})^{10/3}$ 
from neutron star-neutron star binaries with 
angular velocity $\Omega_{orb}$ and chirp mass
${\cal M}=(M_1M_2)^{3/5} /(M_1+M_2)^{1/5}$ (for circular orbits).
The ratio of radio-to-gravitational wave emissions can be evaluated as
\begin{eqnarray}
{{\cal L}_{em}}:
{{\cal L}_{gw}}
\sim (\Omega M)^{2/3} (E_B/M) (M/m)^2 < 1,
\label{EQN_EST}
\end{eqnarray}
e.g., when $E_B/M\sim 10^{-6}$ for the relative
energy in the magnetic field and $M/m\le 10^2$.

The fluence $F_{GW}$ in gravitational radiation may be appreciable, provided
the torus is long-lived in a state of suspended accretion for the
duraction of the burst. The suspended accretion state is described 
by balance of torque and energy: 
\begin{eqnarray}
\left\{
\begin{array}{rl}
\tau_+         &= \tau_-       +\tau_{rad},\\
\Omega_+\tau_+ &=\Omega_-\tau_-+\Omega\tau_{rad}+P_{d},
\end{array}
\right.
\label{EQN_B}
\end{eqnarray}
where $P_{d}$ denotes dissipation,
$\Omega\approx\Omega_K$ a mean orbital angular frequency and
$\tau_-=A^2f_w^2\Omega_-$ denotes the torque on the outer face of the torus.
The net magnetic flux $2\pi A$ supported by the torus will
partially connect to the black hole and to infinity, 
respectively, with
fractions $f_H$ and $f_w$.
Thus, $A\approx ab <B_\theta>$
in terms of the average poloidal component $B_\theta$ in the torus.
Generally, $f_H+f_w={1}/{2}~-~1$ with
$f_H\propto (M/a)^2$ for a slender torus. A
remainder $1-f_H-f_w$ is inactive in closed field-lines 
(with Dirichlet-Dirichlet boundary conditions)
between the inner light surface and the outer light cylinder, 
attached to each face as toroidal ``bags." 
Note that for small differential rotation, we have
$(\Omega_K\tau_{rad}+P_{d})/{\Omega_K\tau_{rad}}
\approx({\Omega_+\tau_+-\Omega_-\tau_-})/\Omega_K({\tau_+-\tau_-})
\approx 2,$
in which limit the efficiency of the radiation is $50\%$.

We shall assume that the coupling between
the two faces is dominated by magnetohydrodynamical
stresses, due to radial components $B_r$ of the magnetic field.
These stresses are dissipative, by Ohmic heating and
magnetic reconnection. This will heat the torus, which brings
about thermal and, possibly, neutrino emission.
By dimensional analysis
\begin{eqnarray}
P_{d}\approx A^2_r(\Omega_+-\Omega_-)^2,~~~
A_r=ah<B_r^2>^{1/2},
\end{eqnarray}
where the second equation denotes the root-mean-square of 
the radial flux averaged over
the interface between the two faces with contact area $2\pi a h$.
While the angular momentum transport in the shear layer about this
interface is mediated by $<B_r^2>^{1/2}$,
the angular momentum transport from the black hole to
the torus is mediated by the average $<B_\theta>$. The first
comprises the spectral density average over all azimuthal
quantum numbers $m$, whereas the second only involves $m=0$.
Indeed, the net flux through the black hole is generated
by the coroting horizon charge $q\approx <B_n>J$ in magnetostatic
equilibrium with the mean external poloidal magnetic field
\citep{wal74,dok87,mvp00b,lee01}.
This averaging process is due to the no-hair theorem. While 
the exact ratio depends on the details of the magnetohydrodynamic 
turbulence in the torus, a conservative estimate is that $A_r/A$ 
is about the square root of the number of azimuthal modes in 
the approximately uniform infrared spectrum, which
should reach up to the first geometrical break at $m=a/b$, i.e.: 
$A_r/A\approx (a/b)^{1/2}$.
Thus, substitution of the first into the second of 
the stationary conditions (\ref{EQN_B}) gives rise to a positive
luminosity
\begin{eqnarray}
\Omega\tau_{rad}\approx\Omega^2A^2\left[3(A_r/A)^2(b/a)-2f_w^2\right]\sim
\Omega^2A^2
\label{EQN_LR}
\end{eqnarray}
in view of $\Omega_-\approx \Omega$, $\Omega_+-\Omega_-\approx (3/2)(b/a)\Omega$
and $f_w<1$. The first
equation of (\ref{EQN_B}) now reduces to
$\Omega/\Omega_H\approx f_H^2/3.$
With $\Omega\approx M^{1/2}/R^{3/2}$ and $f_H\propto (M/a)^2$,
this shows that $R\propto M^{7/5}\Omega_H^{2/5}$, i.e.,
the radius of the torus decreases as the black hole spins-down.
This defines a horizontal branch of the frequency dynamics in
the $\dot{f}(f)-$diagram \citep{mvp00d}. At twice the Keplerian
frequency of the torus, this produces an observed
gravitational wave-frequency of about
\begin{eqnarray}
f_{gw}\sim1-2kHz/(1+z)
\end{eqnarray} 
for canonical GRB values for a black hole-torus 
system at redshift $z$. If the torus is unstable
against breaking up in clumps, or if the torus shows violent
expansions in its mean radius, the gravitational waves will
be episodical, and will correlate with
sub-bursts in long GRBs. 

The above shows that a stationary state of suspended accretion
in the presence of gravitational wave-emissions is facilitated 
by magnetohydrodynamical viscosity. Note that no specific
instability mechanism is identified which is to account for 
the required non-axisymmetric deformations of the torus. It would be of
interest to study this by numerical simulations.

\section{Long GRBs as LIGO/VIRGO sources}

The GRB-afterglow emissions define the isotropic equivalent luminosity of
the black hole in the present black hole-torus model. Given the uniform
magnetization of the horizon, the collateral interaction onto the torus
is of similar intensity per unit opening angle. It follows that
\begin{eqnarray}
{\cal L}_{gw}^{iso}:{\cal L}_{grb}^{iso}\sim 1,
\end{eqnarray}
given that gravitational-wave emission is essentially unbeamed and 
assuming that the larger fraction of the magnetic field-lines
threading the black hole connect to the torus.
Long duration continuous emission with the predicted 
linear chirp is best detected using matched filtering.
Taking into account, therefore, the expected gain
by a factor $\sqrt{n}$ in sensitivity, where $n$ is the number
of cycles in the emission, the effective amplitude
of the gravitational radiation at a distance $D$ satisfies 
\begin{eqnarray}
h_{eff}^{grb}\sim
\left(\frac{M}{D}\right) 
\left(\frac{F_{GW}}{M}\right)^{1/2} (M\Omega)^{-1/2}(1+z)^{-1}
\end{eqnarray}
for a net fluence $F_{GW}$ in gravitational waves.
With a fraction of order unity of
the black hole-luminosity radiated off
in gravitational waves, derived from its spin-energy of
about one-third its total mass, this points towards
GRBs as potentially the most powerful LIGO/VIRGO 
gravitational-wave burst sources in the Universe.
A geometrical beaming factor of $100-200$ gives rise to one
event per year within a distance $D\sim 100$Mpc with
$h_{eff}\sim 10^{-20}$.
Their approximately monochromatic emissions may have 
interesting cosmological applications, assuming no
cosmological evolution in the GRB parameters.

{\bf Acknowledgement.} This research is supported
by NASA Grant 5-7012, an MIT C.E. Reed Fund and a NATO
Collaborative Linkage Grant. The
author thanks S. Kulkarni and R. Weiss for 
stimulating discussions.

\newpage
{\centerline{\bf Figure captions}}
\mbox{}\\
{\bf Figure 1.} {\small 
Schematic view of the magnetosphere of a black hole-torus system
in poloidal cross-section. Shown are the inner and outer
torus magnetosphere, attached to the two faces (middle, right) of the torus
and the open field-lines to infinity supported by the equilibrium
magnetic moment $\mu_H$ of the black hole (left).
The former have Dirichlet-radiative boundary conditions, the latter
have radiative-radiative boundary conditions ($D$ refers to Dirichlet and
$R/\bar{R}$ to radiative outgoing/ingoing). 
The  magnetic moment $\mu_H^e$ arises in equilibrium
with the external magnetic field \citep{mvp00b,lee01} 
provided by the magnetization
$\mu_T$ of the torus. It serves to regulate an even horizon flux about its maximum
at any rotation rate.
A number of inactive field-lines make up inner and outer bags 
attached to the inner and outer faces of the torus with
{\em DD}-boundary conditions, which touch an
inner and outer light surface (dashed curves).
The $D\bar{R}-$ and $DR$-field lines
mediate Maxwell stresses, by equivalence in poloidal topology
to pulsar magnetospheres \citep{gol69,mvp99b}. 
The strength of the Maxwell stresses on the inner and outer faces corresponds, 
respectively, to those on a pulsar with angular velocity 
$-\Omega_{psr}=\Omega_H-\Omega_T$ and $\Omega_{psr}=\Omega_T$, 
where $\Omega_H$ and $\Omega_T$ are the angular
velocities of the black hole and the torus. 
This topological equivalence to pulsars also implies causality for
the black hole
coupling to the inner face. (Reprinted from \cite{mvp00b}.)

\mbox{}\\
{\bf Figure 2.} {\small 
Schematic diagram of the frequency-dynamics trajectory in the $\dot{f}(f)-$diagram
a gravitational wave-signal produced in by a long GRB event from a high-angular
momentum black hole-torus system. The gravitational waves are emitted by the torus 
in suspended accretion. The linear size of the system decreases as the rotational
energy of the black hole is radiated away. This produces an approximately 
horizontal branch on which $\dot{f}\propto f/T$ (right to dashed line) of
the twice-Keplerian frequency 1-2kHz/($1+z$) of the torus around a 10$M_\odot$ 
rapidly spinning black hole at redshift $z$.
When the spin of the black hole falls
below a critical value, suspended ceases and the torus begins
to hyperaccrete onto the black hole. This final phase may be messy,
whereby accretion may excite quasi-normal mode ringing (QNR) of the horizon
\citep{pap01}.
In case of neutron star-black hole coalescence, there exists
a well-defined precursor signal, wherein $\dot{f}\propto f^{11/3}$
(left to dashed line). For hypernovae, precursor 
gravitational wave-emissions are unknown. 
}

\newpage
\mbox{}\\
\vskip2in
\begin{center}
\plotone{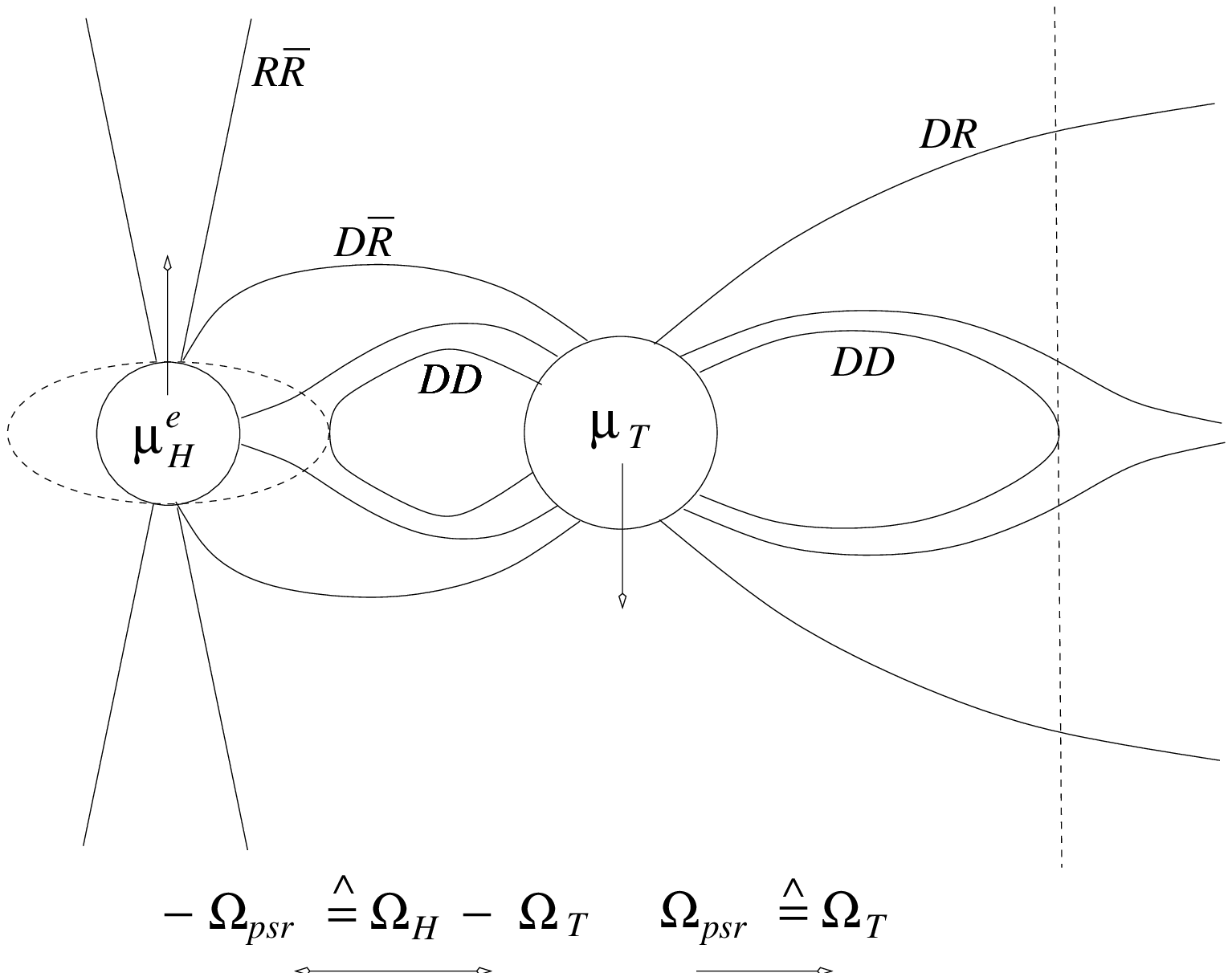}
\end{center}
%\vskip1in
{\sc FIGURE 1}
\newpage
\mbox{}\\
\vskip1in
\begin{center}
\plotone{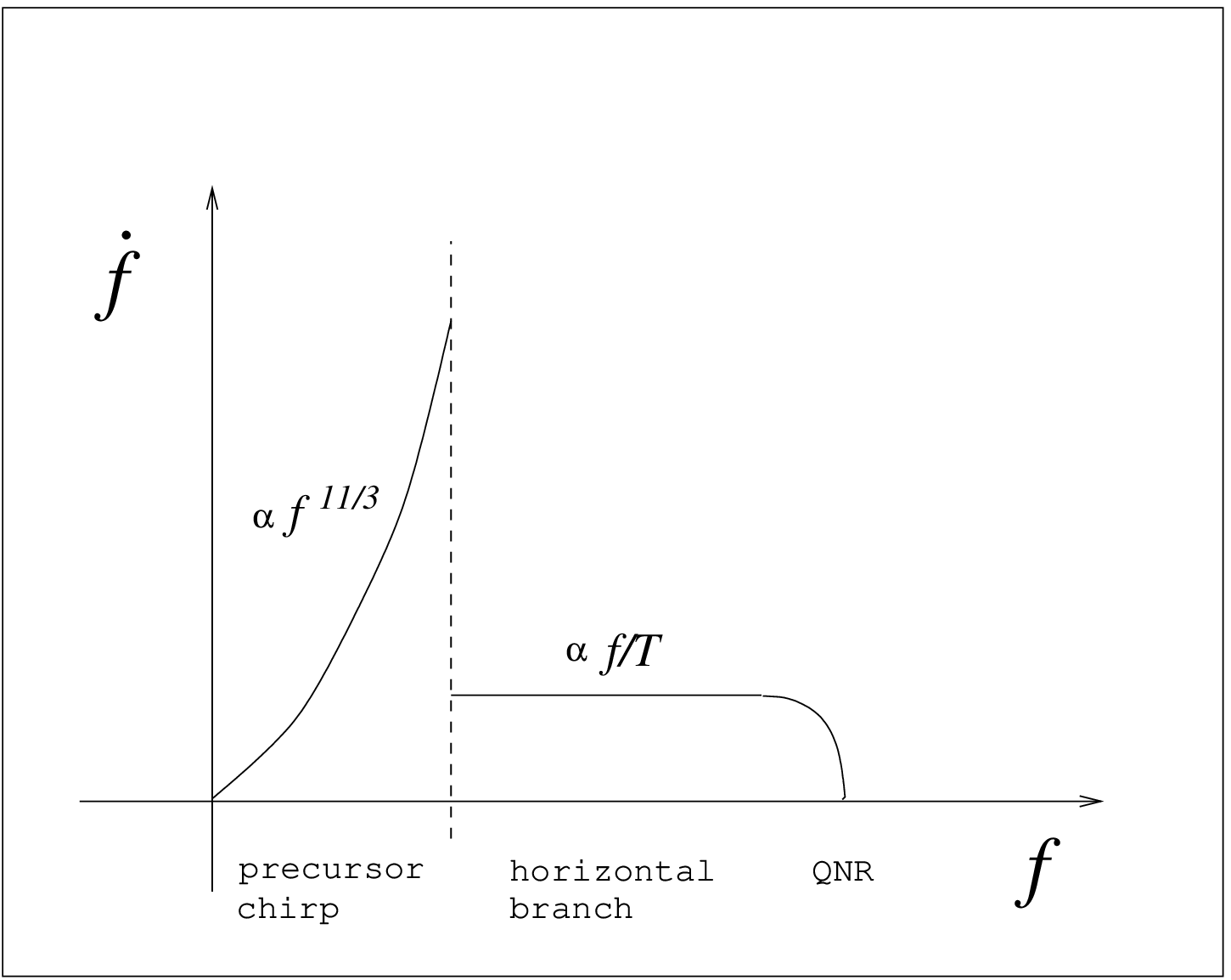}
\end{center}
\vskip1in
{\sc FIGURE 2}

\end{document}